\documentclass[11pt,letterpaper]{article}
\pdfoutput=1

\usepackage[utf8]{inputenc}
\usepackage[T1]{fontenc}
\usepackage{microtype}  
\usepackage[letterpaper,margin=1in]{geometry}
\usepackage{graphicx}
\usepackage{cite}
\usepackage{amsmath,amssymb}
\usepackage{booktabs}   
\usepackage{multirow}   
\usepackage{authblk}    
\usepackage{orcidlink}  
\usepackage[format=plain,labelsep=period,labelfont=bf]{caption}
\usepackage{titlesec}
\titleformat{\section}{\Large\scshape}{\thesection}{0.7em}{}
\titleformat{\subsection}{\large\scshape}{\thesubsection}{0.6em}{}
\titleformat{\subsubsection}{\normalsize\bfseries\itshape}{\thesubsubsection}{0.5em}{}
\usepackage{hyperref}
\hypersetup{hidelinks}

\usepackage{array}  
\newcommand{\captiongap}[1][5.5pt]{\nointerlineskip\vskip #1}
\newcounter{tbox}
\renewcommand{\thetbox}{\arabic{tbox}}

\begin{document}

\title{Perspectives on inverse design for AI magnonics}

\author[1]{Franz Vilsmeier\,\orcidlink{0000-0002-5648-3051}\footnote{franz.vilsmeier@univie.ac.at}}
\author[1]{Florian Bruckner\,\orcidlink{0000-0001-7778-6855}}
\author[1]{Claas Abert\,\orcidlink{0000-0002-4999-0311}}
\author[1]{Dieter Suess\,\orcidlink{0000-0001-5453-9974}}
\author[1]{Andrii V. Chumak\,\orcidlink{0000-0001-5515-0848}\footnote{andrii.chumak@univie.ac.at}}
\affil[1]{\textit{Faculty of Physics, University of Vienna, Vienna, Austria}}

\date{\today}

\maketitle

\begin{abstract}
\noindent Inverse design---specifying a desired functionality and letting a computational algorithm search for a structure that meets it---has emerged as a powerful paradigm for magnonic device engineering.
In this article, we survey the rapidly growing field of inverse-design magnonics, organising it along two axes: the design variables (topology, material parameters, and magnetic field landscape) and the algorithmic toolbox (gradient-free, gradient-based, and neural-network-based methods) together with the differentiable micromagnetic solvers that enable them.
We then identify open frontiers that we consider most promising for the next phase of the field: sensitivity analysis and robust design to bridge the gap between simulation and experiment; input shaping and transducer optimisation; the incorporation of nonlinear spin-wave effects as an explicit design resource; spatially structured amplification; self-adapting media and machine-learning-based design; and the long-term vision of a universal, reconfigurable magnonic platform.
We argue that magnonics and artificial intelligence are converging from two directions---machine-learning tools for designing magnonic devices, and magnonic devices as hardware for neuromorphic computation---and propose the term \emph{AI magnonics} to describe this emerging paradigm.
\end{abstract}

\bigskip

\section{Introduction}
\label{sec:introduction}

Spin waves---the collective excitations of ordered magnetic materials---and their quanta, magnons, have emerged as promising carriers for wave-based information processing~\cite{Chumak2015, Barman2021, Chumak2022, Flebus2024}.
Operating at frequencies from the GHz to the THz range and exhibiting wavelengths down to the nanometre scale, they combine intrinsic nonlinearity with low energy dissipation and offer a route to on-chip computing without charge flow.
The conventional approach to designing magnonic devices---propose a geometry based on physical intuition, simulate its spin-wave response, and refine iteratively---has produced notable results, from nanoscale directional couplers~\cite{Wang2018Coupler, Wang2020NatElec} and majority gates~\cite{Talmelli2020} to all-magnonic repeaters~\cite{Wang2024Repeater}, cascaded nonlinear magnonic circuits~\cite{Guo2026Neurons, Breitbach2026}, and magnon-scattering reservoirs~\cite{Koerber2023}.
However, magnonic devices involve many coupled degrees of freedom, and the underlying physics is sufficiently complex that many high-performing configurations lie beyond the reach of manual parameter studies.
As the field moves toward exploiting nonlinear spin-wave dynamics for neuromorphic and unconventional computing, this design complexity grows further.
How to systematically design structures that exploit this rich physics remains a central open challenge.

\begin{figure}[h]
	\centering
	\includegraphics[width=0.67\textwidth]{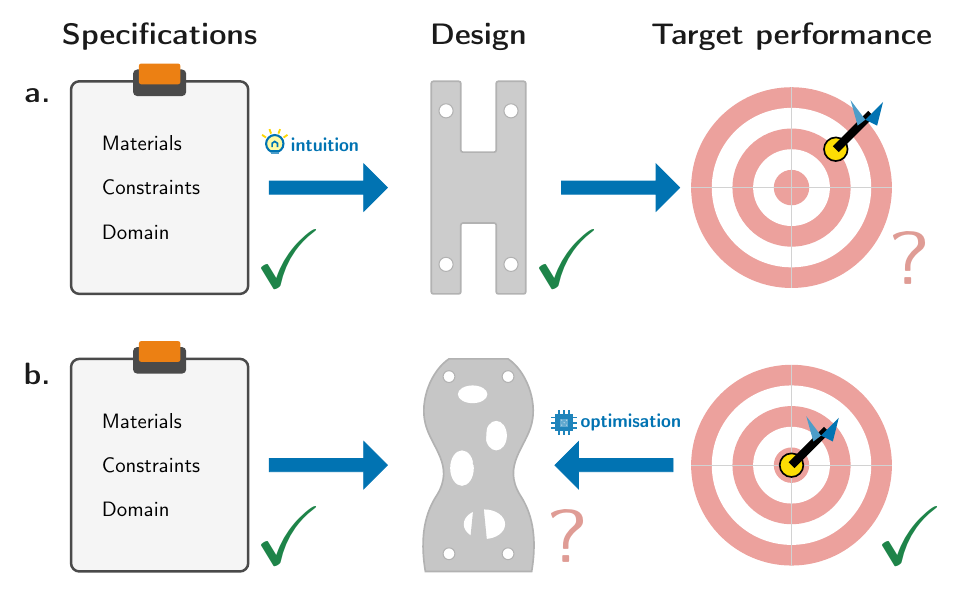}
	\caption{\textbf{Inverse design as a paradigm shift in magnonics.}
		Both workflows share the same three elements---a set of specifications, a device geometry, and a target performance (shown as a dartboard).
		(a)~In direct design, a researcher converts the specifications into a geometry guided by physical intuition, and the resulting performance is only evaluated afterwards, with no guarantee that it meets the target (the dart may miss).
		(b)~In inverse design, the target performance is instead prescribed up front and an optimisation algorithm works \emph{backward} to the device geometry (note the reversed arrow), searching a defined design space for a structure that meets the target, often a non-intuitive one.
		Check marks and question marks indicate which stages are prescribed from the outset and which are only known after evaluation.}
	\label{fig:overview}
\end{figure}

Inverse design addresses this bottleneck: the desired device functionality is specified as an objective function, and a computational algorithm searches automatically for a structure that meets it.
Fig.~\ref{fig:overview} illustrates the paradigm shift from direct to inverse design.
Crucially, by navigating a high-dimensional design space that defies physical intuition, inverse design can identify non-intuitive solutions that a manual parameter study is unlikely to reach.
Outperforming a hand-designed device is not guaranteed, but it happens---topology-optimised structural parts are routinely 20--40\,\% lighter than intuition-based designs~\cite{Aage2017}, an evolved satellite antenna beat its hand-designed counterpart on a NASA mission~\cite{Hornby2011}, and an inverse-designed photonic demultiplexer is two orders of magnitude smaller than conventional ones~\cite{Piggott2015}.
These cases come from a wider tradition. Inverse design is an instance of an inverse problem---inferring unknown causes from desired effects---a concept familiar from fields as diverse as medical imaging and geophysics~\cite{Tarantola2005}.
Computational approaches to inverse design have been applied in structural mechanics~\cite{Bendsoe1988, SigmundMaute2013, Aage2017}, photonics~\cite{Jensen2011, Hughes2018, Piggott2015, Molesky2018}, phononics and acoustics~\cite{Sigmund2003, Dong2024Phononic}, and electromagnetic device engineering~\cite{Bruckner2017,Lucchini2022}.
Among these, photonics represents the most mature ecosystem.
Magnonics, however, offers a richer set of design variables: beyond geometry, the saturation magnetisation $M_\mathrm{s}$, the anisotropy constant $K_\mathrm{u}$, and the Gilbert damping $\alpha$ can in principle be tuned independently, and a reconfigurable external magnetic field landscape provides an additional design variable with no photonic analogue.
Furthermore, spin-wave propagation is inherently nonreciprocal and governed by the nonlinear Landau--Lifshitz--Gilbert (LLG) equation, properties that arise naturally and without the need for special media.

Inverse-design magnonics emerged in 2021 with two independent foundational works.
Wang~\emph{et~al.}~\cite{Wang2021} introduced binary-pixel topology optimisation to the field and demonstrated inverse-designed spin-wave routers and demultiplexers in simulation.
Papp~\emph{et~al.}~\cite{Papp2021} developed a spin-wave neural network in which nanomagnets placed on top of a continuous magnetic film serve as trainable scattering elements, exploiting nonlinear spin-wave interference for pattern recognition---specifically, the recognition of spoken vowels.
The field has since advanced rapidly: magnonic filters optimised via particle swarm optimisation (PSO)~\cite{Yan2022}, level-set topology optimisation~\cite{Voronov2025}, temporal pulse design~\cite{Casulleras2023}, field-free inverse-designed demultiplexers~\cite{Chen2025VCMA}, and dedicated differentiable micromagnetic solvers that share their computational paradigm with neural network training~\cite{Abert2025} have been reported.
Inverse-designed spin-wave lenses~\cite{Kiechle2022}, a universal inverse-design RF device~\cite{Zenbaa2025NatElec, Wu2025}, and inverse-designed logic gates~\cite{Zenbaa2025SciAdv} have been realised experimentally---all within four years of the founding works.
Table~\ref{tab:demonstrations} summarises the inverse-design magnonic demonstrations reported to date.

\begin{table*}[h]
	\caption{\textbf{Summary of inverse-design magnonic demonstrations.} Listed are studies on inverse design in magnonics, published up to July 2026. For each, we give the optimised design variable, the algorithm employed, the target functionality, whether the demonstration was simulation-based (S) or experimental (E), and its key contribution.}
	\captiongap
	\centering
	\footnotesize
	\setlength{\tabcolsep}{2pt}
	\hyphenpenalty=10000\exhyphenpenalty=10000  
	\begin{tabular}{@{}c l l l l c >{\raggedright\arraybackslash}p{3.9cm}@{}}
		\toprule
		\textbf{Year} & \textbf{Reference} & \textbf{Design variable} & \textbf{Algorithm} & \textbf{Objective} & \textbf{S/E} & \textbf{Key contribution} \\
		\midrule
		2021 & Wang~\emph{et al.}~\cite{Wang2021} & Geometry (binary) & DBS & Routing, demux & S & Founding work; topo.\ opt. \\
		2021 & Papp~\emph{et al.}~\cite{Papp2021}$^{\dagger}$ & Nanomagnet states & Backprop & Pattern recog. & S & Founding work; neural network \\
		2022 & Kiechle~\emph{et al.}~\cite{Kiechle2022}$^{\dagger}$ & Material params. & Backprop & SW lens & E & Inverse-designed SW lens \\
		2022 & Yan~\emph{et al.}~\cite{Yan2022} & Magnonic crystal & PSO & Bandpass filter & S & PSO-optimised filter \\
		2023 & Casulleras~\emph{et al.}~\cite{Casulleras2023} & Input pulse & Analytical & Target pulse & S & Temporal inverse design \\
		2025 & Zenbaa~\emph{et al.}~\cite{Zenbaa2025NatElec} & Field landscape & DS, GA & Notch filter, demux & E & Simulation-free, in-situ design \\
		2025 & Zenbaa~\emph{et al.}~\cite{Zenbaa2025SciAdv} & Field landscape & DS & Logic gates & E & Boolean logic \\
		2025 & Voronov~\emph{et al.}~\cite{Voronov2025}$^{\dagger}$ & Geometry (level-set) & Adjoint & Hysteresis, demux & S & Level-set topo.\ opt. \\
		2025 & Chen~\emph{et al.}~\cite{Chen2025VCMA} & Geometry (binary) & Monte Carlo & Demux, mux & S & Field-free (VCMA bias) \\
		2025 & Abert~\emph{et al.}~\cite{Abert2025}$^{\dagger}$ & --- & Diff.\ solver & General-purpose & S & Adjoint-based solver \\
		\bottomrule
	\end{tabular}\\[1pt]
	\parbox{\textwidth}{\scriptsize\linespread{0.8}\selectfont\textbf{DBS}: direct binary search; \textbf{GA}: genetic algorithm; \textbf{PSO}: particle swarm optimisation; \textbf{DS}: direct search; \textbf{Topo.\ opt.}: topology optimisation; \textbf{VCMA}: voltage-controlled magnetic anisotropy; \textbf{SW}: spin wave; \textbf{demux}: demultiplexer; \textbf{mux}: multiplexer; \textbf{S}: simulation; \textbf{E}: experimental. $^{\dagger}$Open-source optimisation framework used.}
	\label{tab:demonstrations}
\end{table*}

In this article, we survey the current state of the art and, based on our experience in the field, identify what we consider the most promising open frontiers---from input shaping and nonlinear design to amplification and self-adapting systems---culminating in the long-term vision of a universal, reconfigurable magnonic platform.
Table~\ref{tab:overview} provides an overview of the content of this article. Box~\ref{box:terminology} distinguishes terms that are most easily grouped together.

\begin{table*}
\caption{\textbf{Overview of inverse-design magnonics literature.}
Papers are grouped by the section in which they are primarily discussed and organised by class;
a paper may appear in more than one class if it contributes to multiple topics.}
\captiongap[4pt]
\centering
\begin{tabular}{l p{7.5cm} p{3.5cm}}
\toprule
\textbf{Class} & \textbf{Description} & \textbf{Key references} \\
\midrule
\multicolumn{3}{l}{\textbf{Sec.~\ref{sec:design-space} --- Design space: topology, material, and field}} \\
\midrule
Topology              & Binary or level-set material distribution optimised for target spin-wave response
                      & \cite{Wang2021, Yan2022, Chen2025VCMA, Voronov2025} \\[2pt]
Material parameters   & Spatially varying $M_\mathrm{s}$, $K_\mathrm{u}$, $\alpha$ via ion irradiation or laser processing
                      & \cite{Kiechle2022, Ruane2017, Kiechle2023, Bensmann2025, Greil2025, Naunheimer2026, Albisetti2016, Levati2023, Giacco2024, Florio2026, Levati2025, Riddiford2025, Dutta2015, Rana2019, Vogel2015, Kuznetsov2025SciAdv, Kuznetsov2025AdvMat} \\[2pt]
Field landscape       & Reconfigurable external or stray-field patterns steering spin-wave propagation
                      & \cite{Papp2021, Zenbaa2025NatElec, Zenbaa2025SciAdv, Qin2022, Li2022ASI, Sultana2025, Baumgaertl2023, Nizet2025, Wagner2016, Ma2015, Cocconcelli2024, Cocconcelli2025,Cocconcelli2026} \\[2pt]
\midrule
\multicolumn{3}{l}{\textbf{Sec.~\ref{sec:algorithms} --- Algorithmic toolbox}} \\
\midrule
Gradient-free         & Genetic algorithms, PSO, direct search, analytical inversion; no derivative required
                      & \cite{Wang2021, Yan2022, Chen2025VCMA, Zenbaa2025NatElec, Zenbaa2025SciAdv, Casulleras2023} \\[2pt]
Gradient-based        & Adjoint method, backpropagation through differentiable solvers; scales to thousands of parameters
                      & \cite{Voronov2025, Papp2021, Kiechle2022} \\[2pt]
Neural-network-based  & Neural networks as surrogate models, inverse-mapping or generative tools (no magnonic demonstration yet)
                      & \cite{Peurifoy2018, Deng2021, LiuTandem2018, Yeung2023, MaLi2020, Hegde2020, JiangFan2019, Sajedian2019, Ma2021DeepLearning} \\[2pt]
Differentiable solvers& Auto-differentiable micromagnetic frameworks enabling backpropagation; AI-assisted generation of scientific solver code
                      & \cite{Papp2021SpinTorch, Bruckner2023, Abert2025, RomeraParedes2024FunSearch, Novikov2025AlphaEvolve, Li2025CodePDE, Lu2026AIScientist, OBrien2025} \\[2pt]
\midrule
\multicolumn{3}{l}{\textbf{Sec.~\ref{sec:perspectives} --- Perspectives and outlook}} \\
\midrule
Robust design         & Sensitivity analysis and fabrication-aware optimisation (first magnonic checks reported; systematic study still absent)
                      & \cite{Wang2021, Papp2021, Kiechle2022, Wang2019, Piggott2017, Schubert2022} \\[2pt]
Input shaping         & Optimised excitation waveforms, antenna geometry, and transducer efficiency
                      & \cite{Casulleras2023, Gruszecki2016, Koerner2017, Wartelle2023, Temdie2024, Temdie2023, Wagle2026, Connelly2021, Erdelyi2025, Bruckner2025Transducer} \\[2pt]
Nonlinear design      & Inverse design exploiting nonlinear spin-wave dynamics
                      & \cite{Wang2021, Papp2021, Zenbaa2025SciAdv, Wang2023Nonlinear, Ge2025, Wang2024Repeater, Guo2026Neurons, Szulc2026, Breitbach2026, Guo2026RNG} \\[2pt]
Amplification         & Spin-torque and parametric amplification as an active design variable
                      & \cite{Merbouche2024, Nikolaev2025, Heinz2022, Li2026} \\[2pt]
Self-adapting systems & Reconfigurable media, neuromorphic hardware, and reservoir computing
                      & \cite{Papp2021, Guo2026Neurons, Breitbach2026, Fripp2026Neurons, Fripp2026Adder, Koerber2023, Nakane2018, Watt2020, Grollier2020, Baumgaertl2023, Nizet2025, Zenbaa2025NatElec, Zenbaa2025SciAdv, Dutta2015, Rana2019, Husain2024, Husain2025npj, Husain2025Newton} \\
\bottomrule
\end{tabular}
\label{tab:overview}
\end{table*}

\begin{table*}[tp]
{
\setlength{\fboxrule}{0.797pt}  
\fbox{\parbox{\dimexpr\columnwidth-2\fboxsep-2\fboxrule\relax}{%
\footnotesize   
\refstepcounter{tbox}\label{box:terminology}
\textbf{Box~\thetbox. Terminology used in this article.}
The terms below rest on substantially different principles but are easily grouped together. The list is not a taxonomy of the field.\\[6pt]
\hyphenpenalty=10000 \exhyphenpenalty=10000
\begin{tabular}{@{}p{0.28\columnwidth} p{0.66\columnwidth}@{}}
\raggedright\textbf{Term} & \raggedright\textbf{Meaning} \tabularnewline
\midrule
\raggedright Inverse design
  & \raggedright A target is specified as an objective function and an optimiser searches the design space for a structure that meets it \tabularnewline[2pt]
\raggedright Neural-network surrogate (Sec.~\ref{sec:ml-approaches})
  & \raggedright A network trained on micromagnetic data replaces the solver, its differentiability enabling gradient-based inverse search \tabularnewline[2pt]
\raggedright Generative model (Sec.~\ref{sec:ml-approaches})
  & \raggedright A network maps a target specification directly to a candidate structure \tabularnewline[2pt]
\raggedright Reinforcement learning (Sec.~\ref{sec:ml-design})
  & \raggedright An agent learns a design policy through direct interaction with the simulation \tabularnewline[2pt]
\raggedright Neuromorphic magnonic computation (Sec.~\ref{sec:self-adapting-media})
  & \raggedright The spin-wave medium performs the computation itself, its response trained rather than programmed \tabularnewline[2pt]
\raggedright Physical reservoir computing (Sec.~\ref{sec:self-adapting-media})
  & \raggedright The intrinsic nonlinearity is exploited without optimising the device structure; only the readout is trained \tabularnewline[2pt]
\raggedright Self-adapting medium (Sec.~\ref{sec:self-adapting})
  & \raggedright A fabricated device whose structure reconfigures in response to its own spin waves \tabularnewline
\end{tabular}%
}}}
\end{table*}

\section{The design space: topology, material, and field}
\label{sec:design-space}

\begin{figure}[h!]
	\centering
	\includegraphics[width=\textwidth]{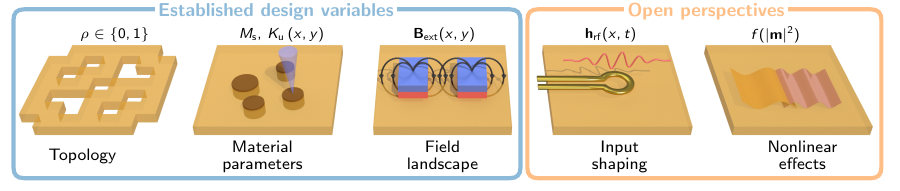}
	\caption{\textbf{Design variables in inverse-design magnonics.}
		From left to right: topology (binary or level-set material distribution),
		material parameters (spatially varying $M_\mathrm{s}$, $K_\mathrm{u}$, $\alpha$),
		and magnetic field landscape (reconfigurable stray-field or current-loop patterns)
		constitute the three established design variables.
		Input shaping (excitation waveforms and antenna geometry)
		and nonlinear effects (frequency shifts or multi-magnon interactions as a design resource)
		are discussed as open perspectives in Sec.~\ref{sec:perspectives}.
	}
	\label{fig:dofs}
\end{figure}

This section surveys what can be optimised in an inverse-design magnonic device; Sec.~\ref{sec:algorithms} then addresses how.
Five design variables can be distinguished: three are established and discussed here, while two further ones---input shaping and nonlinear effects---carry substantial untapped potential and are treated as open perspectives in Sec.~\ref{sec:perspectives}.
These design variables differ fundamentally in their character: some produce permanent modifications and are fixed once fabricated, while others are reconfigurable and can be adapted in situ.
They also differ in their parameterisation: topology is inherently discrete (material present or absent), whereas material parameters and field landscapes are continuously tunable.
Fig.~\ref{fig:dofs} provides a schematic overview of all five design variables.
Each of these design variables comprises many individual degrees of freedom---the pixels of a topology or the cells of a field landscape---whose number sets the dimensionality of the optimisation (Sec.~\ref{sec:open-questions}).
Because a fabricated device inevitably departs from its optimised design, how much performance is lost to such deviations is a question for each design variable; Sec.~\ref{sec:robust-design} returns to it as one of the open frontiers.

\subsection{Topology and geometry}
\label{sec:topology}

The most direct analogue to photonic inverse design is topology optimisation: determining where magnetic material is present or absent~\cite{Jensen2011, Piggott2015}.
Two approaches have been explored in magnonics.
Binary pixel grids~\cite{Wang2021, Yan2022, Chen2025VCMA} discretise the design region into cells that are either filled or empty, producing staircase-like boundaries.
Level-set methods~\cite{Voronov2025} instead represent the device boundary as a smooth, continuous surface, yielding geometries more compatible with standard lithographic fabrication.
Using these representations, inverse-designed spin-wave routers, demultiplexers, and bandpass filters have been demonstrated in simulation~\cite{Wang2021, Yan2022, Voronov2025}.
Chen~\emph{et~al.}~\cite{Chen2025VCMA} further showed that voltage-controlled magnetic anisotropy (VCMA) can replace the external bias field, enabling field-free topology optimisation and cascading of inverse-designed components into a magnonic half adder.
The optimisation algorithms employed are discussed in Sec.~\ref{sec:algorithms}.
Topology optimisation is inherently non-reconfigurable, and any fabricated device would be limited by the minimum feature size that can be patterned in the magnetic film. Electron-beam lithography reaches sub-20\,nm features in high-resolution processes, and YIG conduits as well as permalloy nanostripes have been realised with lateral dimensions down to 50\,nm~\cite{ Baumgaertl2023, Heinz2020}.
To date, however, all topology-optimised magnonic devices remain at the simulation stage; no experimental realisation has yet been demonstrated.

\subsection{Material parameters}
\label{sec:material-parameters}

Rather than removing material, one can spatially vary the magnetic properties---$M_\mathrm{s}$, $K_\mathrm{u}$, or $\alpha$---to create an effective refractive-index landscape for spin waves. These modifications are generally permanent.
A rich fabrication toolbox is available for this purpose.
Ion irradiation---using Ga$^+$ focused ion beam (FIB), He$^+$, or Si implantation---tunes $M_\mathrm{s}$ and $K_\mathrm{u}$ through strain-induced magnetoelastic effects, with $\alpha$ adjustable almost independently using He$^+$ ions~\cite{Ruane2017, Kiechle2023, Greil2025, Naunheimer2026, Bensmann2025}.
These techniques have enabled gradient-index spin-wave lenses and prisms~\cite{Kiechle2023}, dispersion-tuned waveguides for large magnonic networks~\cite{Bensmann2025}, and magnetoelastic spin-wave routing~\cite{Greil2025, Naunheimer2026}.
Laser-based phase nanoengineering~\cite{Albisetti2016, Levati2023} provides complementary capabilities, exploiting local 405\,nm irradiation of yttrium iron garnet (YIG) to either crystallise amorphous films or amorphise crystalline films, both defining ferrimagnetic waveguide channels within a paramagnetic matrix~\cite{Giacco2024, Florio2026}.
Controlled melting and recrystallisation of crystalline YIG instead creates a strained phase with up to 30-fold enhanced perpendicular magnetic anisotropy, enabling direct-write three-dimensional magnonic crystals~\cite{Levati2025}, while direct-write laser annealing of metallic magnetic films produces continuous two-dimensional gradients in $M_\mathrm{s}$, $K_\mathrm{u}$, and interlayer exchange coupling~\cite{Riddiford2025}.
Reconfigurable alternatives include VCMA, which tunes $K_\mathrm{u}$ electrically via magneto-electric coupling~\cite{Dutta2015, Rana2019}, and thermoplasmonic modulation of $M_\mathrm{s}$~\cite{Vogel2015, Kuznetsov2025SciAdv, Kuznetsov2025AdvMat}.

Material parameters were first employed as a design variable by Kiechle~\emph{et~al.}~\cite{Kiechle2022}, who predicted the ion-irradiation pattern for a target spin-wave lens and validated the result experimentally via Ga$^+$ FIB on YIG.

\subsection{Magnetic field landscape}
\label{sec:field-landscape}

A third design variable leaves the magnetic medium untouched and instead shapes its magnetic environment.
A spatially inhomogeneous bias field modifies the local spin-wave dispersion and can thereby steer, filter, or redirect propagating spin waves.
Several mechanisms are available at different length scales.
At the millimetre scale, DC current-loop grids placed above a continuous YIG film provide a fully software-defined field landscape~\cite{Zenbaa2025NatElec, Zenbaa2025SciAdv}.
At the micro- to nanoscale, patterned magnetic overlayers offer finer control: CoFeB nanostripes on YIG define spin-wave nanochannels~\cite{Qin2022}, and artificial spin ice provides writable stray-field landscapes~\cite{Li2022ASI, Sultana2025}.
Nanomagnets whose state can be reversed by propagating magnons enable non-volatile magnon memory~\cite{Baumgaertl2023, Nizet2025}.
Domain walls~\cite{Wagner2016} and skyrmion arrays~\cite{Ma2015} constitute reconfigurable spin-wave nanochannels and dynamic magnonic crystals, while MEMS-integrated permanent micromagnets that can be physically repositioned relative to the waveguide offer a further path to miniaturised field control~\cite{Cocconcelli2024, Cocconcelli2025}. A monolithic extension of this approach integrates the magnonic waveguide directly onto a suspended piezoelectric bridge, where voltage-induced strain controls spin-wave propagation via magnetoelastic coupling~\cite{Cocconcelli2026}.

Papp~\emph{et~al.}~\cite{Papp2021} used nanomagnet stray fields as trainable design parameters in a spin-wave neural network, optimising the magnetisation states to perform pattern recognition---one of the two foundational demonstrations of inverse-design magnonics and the first to exploit the field landscape as a design variable.
Zenbaa~\emph{et~al.}~\cite{Zenbaa2025NatElec, Zenbaa2025SciAdv} demonstrated the first experimental inverse-design magnonic devices using this design variable: with a $7\times 7$ current-loop grid, they optimised notch filters, demultiplexers, and logic gates directly on the physical hardware, without requiring simulation.
This in-situ approach provides inherent robustness to fabrication imperfections, since the field pattern can be re-optimised to compensate for manufacturing tolerances---a significant advantage over static structural designs that must bridge the simulation-to-experiment gap.

\section{The algorithmic toolbox}
\label{sec:algorithms}

\begin{figure}[h]
	\centering
	\includegraphics[width=\textwidth]{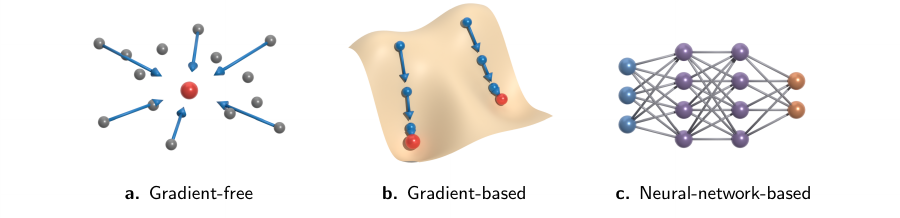}
	\caption{\textbf{The algorithmic toolbox of inverse-design magnonics.}
		(a)~Gradient-free methods evaluate a population of candidate designs (grey) directly, iteratively converging on the best candidate found (red) without using any derivative information---for example direct binary search~\cite{Wang2021}, genetic algorithms, and direct search~\cite{Zenbaa2025NatElec}.
		(b)~Gradient-based methods follow the gradient of the objective downhill across a non-convex landscape, converging to a minimum (red) while risking entrapment in local minima; examples include the adjoint method~\cite{Voronov2025} and backpropagation through a differentiable solver~\cite{Papp2021, Kiechle2022, Abert2025}.
		(c)~Neural-network-based methods employ a trained network---as a differentiable surrogate or a generative model---whose forward and backward passes accelerate or replace the forward simulation; these are well established in photonics but not yet demonstrated in magnonics.}
	\label{fig:algorithms}
\end{figure}

Having established what can be optimised, we now turn to how. The design space is typically high-dimensional and the objective landscape non-convex.
This non-convexity is intrinsic to magnonics: the nonlinearity of the underlying LLG dynamics produces a rugged landscape with many local optima, so the distinction between local and global optimisation is not a technicality but a central design consideration.
Gradient-based methods are \emph{local} optimisers---they follow the objective downhill to the nearest minimum---and are far more efficient when applicable, scaling to thousands of parameters, yet they offer no guarantee of reaching the global optimum and may settle into suboptimal designs.
Gradient-free methods explore the landscape more \emph{globally} and can escape local minima, at the cost of an evaluation count that grows rapidly with dimensionality.
Gradient-based methods require a differentiable forward model; gradient-free methods provide an alternative when this condition is not met or when the design space is inherently discrete. In addition, Casulleras~\emph{et~al.}~\cite{Casulleras2023} demonstrated that for certain problems---specifically temporal pulse design---the inverse problem can be solved analytically via direct inversion of the forward transfer function, bypassing iterative optimisation entirely.
Neural-network-based approaches have not yet been demonstrated in magnonics and are discussed briefly below; the broader artificial intelligence (AI) frontier is treated as a perspective in Sec.~\ref{sec:perspectives}.
Fig.~\ref{fig:algorithms} gives a schematic overview of the three classes of optimisation algorithm.

\subsection{Optimisation algorithms}
\label{sec:opt-algorithms}

\subsubsection{Gradient-free methods}
\label{sec:gradient-free}

Gradient-free methods explore the design space without computing derivatives, evaluating candidate designs directly and using heuristic rules to guide the search toward better solutions.
Two families have been employed in magnonics.
Single-candidate strategies iteratively modify one design and retain each change that improves the objective---flipping individual pixels (direct binary search~\cite{Wang2021}), altering several pixels at random (Monte Carlo search~\cite{Chen2025VCMA}), or stepping through the parameters one at a time (direct search~\cite{Zenbaa2025NatElec}).
Population-based strategies instead maintain many candidates in parallel: genetic algorithms evolve them through selection, crossover, and mutation~\cite{Zenbaa2025NatElec}, while PSO guides them toward the best designs found so far~\cite{Yan2022}.
These methods are the natural choice for discrete design spaces, such as binary-pixel topology grids, where gradients are not defined, and have been the approach of choice for experimental inverse design in magnonics to date~\cite{Zenbaa2025NatElec, Zenbaa2025SciAdv}.
The principal limitation is scaling: the number of required evaluations grows rapidly with the dimensionality of the design space, restricting gradient-free approaches to design spaces of moderate dimensionality.

\subsubsection{Gradient-based methods}
\label{sec:gradient-based}

For the continuous design variables---material parameters and field landscapes---gradient information is directly available and provides an efficient means of navigating high-dimensional landscapes.
Topology, while inherently discrete, becomes accessible to gradient-based methods through a continuous parameterisation such as the level-set method.
The procedure can be pictured as descending a hilly landscape: from the current position, one takes a step along the locally steepest downhill direction and repeats until the terrain flattens (Fig.~\ref{fig:algorithms}(b)).
At each iteration, the gradient of the objective function with respect to all design parameters provides exactly this direction, and the design is updated accordingly.
The approach scales efficiently to thousands of parameters.
Papp~\emph{et~al.}~\cite{Papp2021, Papp2021SpinTorch} introduced gradient-based inverse design to magnonics using the differentiable solver SpinTorch~(Sec.~\ref{sec:diff-solvers}), optimising nanomagnet states for spin-wave pattern recognition via backpropagation through the time-stepped LLG equation.
Kiechle~\emph{et~al.}~\cite{Kiechle2022} used the same framework to design the ion-irradiation pattern for a spin-wave lens and validated the result experimentally.
Voronov~\emph{et~al.}~\cite{Voronov2025} applied the adjoint method~\cite{Hughes2018} via NeuralMag~(Sec.~\ref{sec:diff-solvers}) to level-set topology optimisation, demonstrating the approach on both the shape-dependent hysteresis of a Stoner--Wohlfarth particle and a magnonic demultiplexer, converting the binary topology problem into a continuous one.
A well-known caveat is convergence to local minima---the descent ends in the nearest valley, which need not be the deepest---a risk that grows as devices are driven into the strongly nonlinear regime (Sec.~\ref{sec:nonlinear-dof}). Multi-start (random-restart) optimisation, basin hopping, continuation methods that ramp up the nonlinearity gradually, and hybrid approaches that alternate gradient descent with heuristic perturbations represent promising directions for mitigating this.
Computationally, these workflows share their infrastructure---backpropagation, loss functions, and optimisers---with neural network training (Sec.~\ref{sec:diff-solvers}), providing a natural bridge to the neural-network-based approaches discussed next.

\subsubsection{Neural-network-based approaches}
\label{sec:ml-approaches}

A qualitatively different strategy is to let AI perform the design task itself, employing a neural network as a surrogate forward model whose differentiability enables gradient-based inverse search~\cite{Peurifoy2018, Deng2021}, or as a generative model that maps a target specification to a candidate design~\cite{Yeung2023}.
Tandem architectures that chain a forward surrogate with an inverse network address the non-uniqueness of the inverse mapping~\cite{LiuTandem2018}.
A complementary hybrid strategy pairs a neural network surrogate with a gradient-free optimiser such as PSO or an evolutionary algorithm, combining global search with fast evaluation~\cite{MaLi2020, Hegde2020}.
All of these approaches are well established in photonics but have not yet been demonstrated in magnonics.
The principal challenges are the cost of generating training data---each sample requires a full micromagnetic simulation---and generalisation beyond the training distribution.
The differentiable solvers discussed in Sec.~\ref{sec:diff-solvers} provide a compatible infrastructure: they operate in the same frameworks (PyTorch, JAX), generate data with gradient information, and can serve as teacher models for neural network surrogates.
Further approaches, including physics-driven generative adversarial networks for topology optimisation~\cite{JiangFan2019} and reinforcement learning~\cite{Sajedian2019}, have shown considerable promise in photonics (see~\cite{Ma2021DeepLearning} for a comprehensive review); we discuss these as an open frontier in Sec.~\ref{sec:perspectives}.

\subsection{Micromagnetic simulation as the inverse-design engine}
\label{sec:diff-solvers}

The forward model---the simulation that predicts spin-wave behaviour for a given design---is the computational engine of the inverse-design loop.
Standard micromagnetic solvers such as MuMax3 and OOMMF solve the LLG equation accurately but are implemented as standalone tools without automatic differentiation, restricting their use to gradient-free optimisation.
The development of differentiable micromagnetic solvers has therefore been a critical enabling step:
\begin{itemize}
\item \textbf{SpinTorch}~\cite{Papp2021SpinTorch} implements backpropagation through time in PyTorch, originally developed for spin-wave neural networks.
\item \textbf{magnum.np}~\cite{Bruckner2023} provides a general-purpose, GPU-accelerated solver built on PyTorch with automatic differentiation, achieving performance competitive with MuMax3.
\item \textbf{NeuralMag}~\cite{Abert2025} is an open-source nodal finite-difference solver built specifically for inverse design: it implements different strategies for gradient computations in the time domain, including the adjoint method and backpropagation with checkpointing through the diffrax library~\cite{kidger2021on}. Its nodal discretisation is naturally suited to continuous parameterisations such as level-set boundaries and material gradients, and it supports both PyTorch and JAX backends.
\end{itemize}
All three are openly available, which makes them the most direct entry point for groups new to the field.
Simulation time remains the primary constraint on the complexity of inverse-design problems that can be tackled---a single evaluation may take seconds to hours depending on the device size and physics involved, and a full optimisation requires hundreds to thousands of evaluations.
We note that the in-situ approach of Sec.~\ref{sec:field-landscape} requires no micromagnetic simulation at all: the physical device itself provides the spin-wave response, and each evaluation is a direct measurement~\cite{Zenbaa2025NatElec}.
SpinTorch, magnum.np, and NeuralMag all borrow the computational infrastructure of machine learning---automatic differentiation, GPU-accelerated tensor operations, and gradient-based optimisation loops---but apply it to the governing physics rather than to a learned model.
This shared infrastructure makes the magnonic inverse-design ecosystem natively compatible with neural-network-based approaches: the same frameworks, training loops, and data pipelines can serve both physics-based optimisation and the training of neural network surrogates or generative models.

Beyond their role as the forward engine of optimisation, these solvers are themselves software artefacts---and the way scientific software is written is beginning to change.
Large language models (LLMs) and agentic coding tools can increasingly generate, extend, and even discover numerical code: program-search systems have found improved algorithms for open mathematical problems~\cite{RomeraParedes2024FunSearch} and for matrix multiplication~\cite{Novikov2025AlphaEvolve}, LLM-driven frameworks now synthesise and refine partial differential equation solvers~\cite{Li2025CodePDE}, and autonomous agents can carry out end-to-end computational studies from code to manuscript~\cite{Lu2026AIScientist}.
For inverse-design magnonics, which relies on custom differentiable solvers rather than turnkey packages, such tools lower the barrier to implementing new physics terms, porting solvers across frameworks and hardware, and generating performance-critical kernels.
These capabilities carry a clear caveat: AI-generated scientific code can harbour silent errors, and most research code is insufficiently tested, so verification against established solvers and reproducibility safeguards remain essential~\cite{OBrien2025}.
Used judiciously, AI-assisted programming is poised to accelerate development of the very simulation infrastructure on which inverse-design magnonics depends---part of the broader transformation of computational science by AI that underpins the field's tooling.

\section{Perspectives and outlook}
\label{sec:perspectives}

The preceding sections have established the design space, the algorithmic toolbox, and the enabling simulation infrastructure.
We now identify, based on our assessment of the field, the frontiers we consider most promising for the next phase of inverse-design magnonics: five concrete directions, the open questions and fundamental limits that remain, and the long-term vision of a universal magnonic platform.

\subsection{Sensitivity analysis and robust design}
\label{sec:robust-design}

The near-term milestone for the field is the fabrication of the first topology-optimised magnonic device, bridging the simulation results of Wang~\emph{et~al.}~\cite{Wang2021} and Voronov~\emph{et~al.}~\cite{Voronov2025} to experiment.
Inverse-designed devices are optimised for a specific set of parameters; real fabrication inevitably introduces deviations in feature size, material properties ($M_\mathrm{s}$, $K_\mathrm{u}$, $\alpha$), and field homogeneity, while the excitation and detection add antenna misalignment and impedance mismatch, and drift, field calibration and measurement noise enter on the characterisation side.
Quantifying how device performance degrades under such perturbations---from the simulation side by sweeping parameters around the optimum, and from the experimental side by characterising multiple fabricated samples---is essential for establishing whether an inverse-designed structure can be realised at all.
Such checks have already been reported, but only in isolated and largely qualitative form.
In simulation, Wang~\emph{et~al.}~\cite{Wang2021} found their demultiplexer to tolerate finite temperature and a fabrication-induced enlargement of the etched voids, while Papp~\emph{et~al.}~\cite{Papp2021} reported the nanomagnet states to be insensitive to switching errors; experimentally, Kiechle~\emph{et~al.}~\cite{Kiechle2022} traced the reduced quality of their fabricated spin-wave lens to fabrication-induced deviations in the local magnetisation.
A systematic sensitivity analysis---particularly a dedicated experimental tolerance study---remains absent.
In photonics, robust optimisation---which targets worst-case or average performance over a distribution of parameter perturbations---and fabrication-aware regularisation via erosion and dilation filters are well established~\cite{Wang2019, Piggott2017, Schubert2022}.
Magnetic field-landscape inverse design~\cite{Zenbaa2025NatElec} offers built-in robustness through in-situ re-optimisation, but static structural designs must bridge the simulation-to-experiment gap without this safety net.
Re-optimisation is not unconditional, however. Hysteresis in the field-generating elements, noise or drift in the applied field, and temperature changes all shift the response, so a configuration found in situ need not remain optimal.
Establishing fabrication tolerances for each design variable---for example the deviation that still keeps the performance within 1\,dB of its nominal value---and incorporating them into the optimisation loop is an essential next step.

\subsection{Input shaping and excitation}
\label{sec:input-shaping}

Casulleras~\emph{et~al.}~\cite{Casulleras2023} showed that for temporal pulse design the inverse problem can be solved analytically in the linear regime.
We see considerably more potential, however, in the spatial side of the excitation: antenna geometry and transducer design.
The shape of the excitation antenna controls beam collimation, focusing, and diffraction~\cite{Gruszecki2016, Koerner2017, Wartelle2023, Temdie2024, Temdie2023, Wagle2026}, yet all layouts have been designed manually.
Transducer efficiency is a closely related constraint: recent work has established system-level models~\cite{Connelly2021}, scaling rules~\cite{Erdelyi2025}, and micromagnetic approaches to computing the spin-wave resistance~\cite{Bruckner2025Transducer}, all using manual parameter sweeps rather than systematic optimisation.
Applying inverse design to the excitation---from antenna shape to transducer impedance matching---could substantially improve the efficiency of spin-wave injection, which remains one of the key practical bottlenecks in magnonic circuits.

\subsection{Nonlinear effects}
\label{sec:nonlinear-dof}

Nonlinearity has been part of magnonic inverse design from the very beginning: the founding works demonstrated a power-dependent spin-wave switch~\cite{Wang2021} and a spin-wave neural network whose computational power rises sharply in the nonlinear regime~\cite{Papp2021}.
In both cases, however, nonlinearity was exploited rather than designed: the optimisation acted on the structure while the nonlinear physics was taken as given.

More fundamentally, nonlinearity is necessary for entire classes of functionality.
Linear interference, however richly it connects inputs and outputs, acts as a single linear layer on the input amplitudes: it cannot realise linearly non-separable functions such as XOR, let alone the multilayer mappings underlying neuromorphic computation~\cite{Papp2021}.
This limitation assumes information is amplitude-encoded and read linearly. While phase-encoding and nonlinear power detection can produce an XOR-like response in a linear system, this only occurs at the final readout and cannot be cascaded within the device itself.
The field-landscape platform of Sec.~\ref{sec:field-landscape} confirms the role of the medium's nonlinearity experimentally: linear operation yields RF components such as notch filters and demultiplexers~\cite{Zenbaa2025NatElec}, whereas Boolean logic gates and a half-adder required driving the spin waves into the nonlinear regime~\cite{Zenbaa2025SciAdv}.

This added capability has a concrete physical origin.
At large precession angles the static magnetisation is reduced; in the out-of-plane geometry this lowers the demagnetising field and shifts the spin-wave dispersion upward---by more than 2\,GHz in nanoscale YIG waveguides at a precession angle of about 55$^\circ$~\cite{Wang2023Nonlinear, Szulc2026}---while in-plane geometries show the opposite sign~\cite{Lvov1994, Krivosik2010, Lake2022, Bunkov2023}.
Because the reduction of the static magnetisation follows the local wave amplitude, the interference pattern imprints a spatially non-uniform modulation on the medium: the wave reshapes the very landscape through which it propagates.
The reachable space of device behaviours thus grows without any added design parameters, and the excitation amplitude becomes a design knob in its own right.

Magnonics offers a richer nonlinear landscape than photonics: beyond the amplitude-dependent frequency shift, spin waves support three- and four-magnon scattering~\cite{Lvov1994, Serga2010} and parametric amplification (Sec.~\ref{sec:parametric-amp}), compared with the single Kerr nonlinearity of photonics~\cite{Hughes2018}.
For inverse design, the key distinction is between deterministic and stochastic effects.
Deterministic effects---chiefly the frequency shift---are reproducible and fully captured by the LLG equation; they have enabled threshold switching in a directional coupler~\cite{Ge2025}, bistable switching in an all-magnonic repeater~\cite{Wang2024Repeater}, and cascaded magnonic threshold neurons (Sec.~\ref{sec:self-adapting})~\cite{Guo2026Neurons}.
Stochastic effects, by contrast, produce outputs that differ even for identical inputs; in magnonics they typically stem from thermal fluctuations, with multi-magnon scattering and the noise-induced switching between bistable spin-wave states as prominent examples.
Such stochasticity need not be detrimental: precisely this switching provides the entropy source of a magnonic true-random-number generator~\cite{Guo2026RNG}.

The open frontier is to incorporate nonlinearity explicitly into the optimisation loop---designing not just the structure but the nonlinear operating point, for example, by co-optimising the device geometry and the excitation amplitude to target a specific regime.
For deterministic nonlinearities, the forward model remains fully predictive and gradient-based optimisation applies, though not always reliably. A bistable device can settle in either state, so its response depends on the initial condition as well as on the design, and a device operated close to a threshold is by construction most sensitive there. Ramping the drive amplitude gradually allows a single branch to be tracked, and the strategies of Sec.~\ref{sec:gradient-based} become correspondingly more important. A thresholded output requires one further consideration, as the decision itself offers no gradient and must be replaced by a differentiable approximation such as a sigmoid. Stochastic effects may instead require gradient-free or statistical approaches, such as objectives averaged over many realisations. This is a qualitatively different algorithmic challenge with no direct photonic analogue.

\subsection{Amplification}
\label{sec:parametric-amp}

Propagation loss limits the length and complexity of magnonic circuits.
Two amplification mechanisms have recently been demonstrated at the nanoscale: spin-orbit-torque amplification in BiYIG/Pt waveguides~\cite{Merbouche2024}, achieving up to a fivefold intensity increase, and parametric pumping at $2f$ in YIG nanowaveguides at zero bias field~\cite{Nikolaev2025}, with gains exceeding two orders of magnitude---following the earlier parametric generation of spin waves in such nanoconduits~\cite{Heinz2022}.
Four-wave mixing of a propagating pump wave can moreover generate spontaneously oscillating magnon modes, which amplify weak signals with gains of up to 40\,dB in a YIG delay line~\cite{Li2026}.
For inverse design, the key point is that these amplification centres are not scatterers---they do not redirect spin waves---but spatially localisable gain elements whose placement can itself be engineered.
A structured arrangement could amplify selectively along chosen paths or restore the signal amplitude at specific nodes; jointly optimising the passive structure and the gain distribution would then integrate amplification and signal processing into a single element that can be cascaded with subsequent blocks.

\subsection{Self-adapting and learning systems}
\label{sec:self-adapting}

Two frontiers emerge where magnonics meets machine learning: a self-adapting medium whose physical structure reconfigures in response to its own spin waves---and can itself perform neuromorphic computation---and machine-learning-based design tools that learn to propose better structures from data.

\subsubsection{Self-adapting media and neuromorphic hardware}
\label{sec:self-adapting-media}

The key idea is that propagating spin waves can modify the medium itself, creating a feedback loop.
Nanomagnets reversible by propagating magnons~\cite{Baumgaertl2023, Nizet2025} provide non-volatile magnon memory.
Domain walls~\cite{Wagner2016} and skyrmion arrays~\cite{Ma2015} can be displaced by spin-wave momentum transfer to form writable channels.
Nonlinear effects (Sec.~\ref{sec:nonlinear-dof}) allow the spin-wave amplitude to modify the effective field landscape dynamically.
VCMA offers a complementary, electrically gated route to reconfiguration~\cite{Dutta2015, Rana2019}, in which the spin-wave medium is reshaped with the periodic assistance of an applied electric field---a quasi-self-adapting scheme.
Multiferroics are a promising medium in this context: an electrically written polarisation state controls magnon transport and is retained when the voltage is removed~\cite{Husain2024, Husain2025npj, Husain2025Newton}.
Papp~\emph{et~al.}'s~\cite{Papp2021} spin-wave neural network, in which nanomagnet states are optimised offline via backpropagation through the physics, already provides a prototype for trainable magnonic hardware, exploiting nonlinear spin-wave interference for pattern recognition.
Combined with nanomagnets writable by propagating magnons~\cite{Baumgaertl2023, Nizet2025}, such a system could in principle update its configuration through its own signals.
Recent work has extended this neuromorphic direction: Guo~\emph{et~al.}~\cite{Guo2026Neurons} realised cascaded nonlinear magnonic threshold neurons---with programmable weighting, tunable firing thresholds, and self-normalised output---and integrated seven of them into a network performing reconfigurable pattern recognition, while Breitbach~\emph{et~al.}~\cite{Breitbach2026} realised an all-magnonic neuron with thresholded firing, fading memory, and cascaded triggering.
In simulation, Fripp~\emph{et~al.}~\cite{Fripp2026Neurons, Fripp2026Adder} proposed a neuron concept based on nonlinear resonant scattering from a permalloy disk above a YIG film, and extended it to a magnonic full adder by training a linear readout layer on the spin-wave interference patterns of a multi-disk array.
Physical reservoir computing offers a complementary route, exploiting intrinsic magnonic nonlinearity without optimising the device structure.
Garnet films and delay-line resonators served as early spin-wave reservoirs~\cite{Nakane2018, Watt2020}, within the broader neuromorphic-spintronics effort~\cite{Grollier2020}, and K{\"o}rber~\emph{et~al.}~\cite{Koerber2023} showed that three-magnon scattering in a magnetic vortex disk can serve as a reservoir for temporal pattern recognition.
Nonlinear feedback---routing part of the output back to the input---is a further avenue toward increased autonomy of magnonic devices.
Energy efficiency is a further motivation, given the rapidly growing power draw of AI systems. A magnonic half-adder with 100-nm feature size is estimated at 24.6\,aJ per operation, close to the 35.3\,aJ of 7-nm CMOS, but it occupies five times the area. Downscaling to 30\,nm is projected to give 1.96\,aJ on the CMOS area~\cite{Wang2020NatElec}. These numbers count the magnonic domain alone, excluding the transducers (Sec.~\ref{sec:input-shaping}) and the amplifiers that restore the output level for cascading (Sec.~\ref{sec:parametric-amp}).

\subsubsection{Machine-learning-based design}
\label{sec:ml-design}

Neural-network-based design tools that are well established in photonics~\cite{Ma2021DeepLearning} have not yet been transferred to magnonics.
Forward surrogate networks trained on micromagnetic data could accelerate design-space exploration by orders of magnitude compared to full simulation~\cite{Peurifoy2018, Deng2021}, while generative models could propose candidate designs that are subsequently refined by adjoint optimisation~\cite{Yeung2023, JiangFan2019}.
The compatible infrastructure of the differentiable solvers described in Sec.~\ref{sec:diff-solvers} provides a natural on-ramp for this transition.
Optimisation runs also produce structure--response pairs as a by-product; collected in a shared repository, they could serve as training data and accelerate future optimisations.
Reinforcement learning~\cite{Sajedian2019}, which requires no pre-collected training data and learns a design policy through direct interaction with the simulation environment, is a particularly promising yet unexplored direction for magnonics.
A further, more speculative extension concerns the design tools themselves: the AI coding assistants now able to generate and refine numerical solvers (Sec.~\ref{sec:diff-solvers}) could increasingly automate the construction of magnonic design pipelines, so that AI contributes not only the proposed structures but the software that produces them.

\begin{figure}[h]
	\centering
	\includegraphics[width=0.55\textwidth]{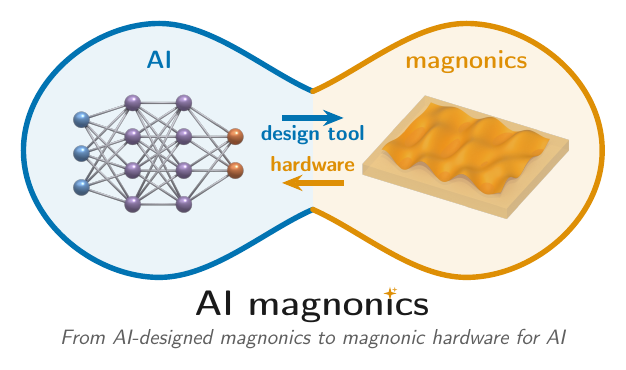}
	\caption{\textbf{AI magnonics: the convergence of inverse design and neuromorphic magnonics.}
		In the emerging \emph{AI magnonics} paradigm, the two fields converge from two directions---machine-learning tools design the device, while the magnonic spin-wave medium itself performs neuromorphic computation, depicted here as two interpenetrating lobes---an artificial neural network and a propagating spin-wave medium---that meet and fuse at the point where the two disciplines converge (\emph{from AI-designed magnonics to magnonic hardware for AI}).}
	\label{fig:ai-magnonics}
\end{figure}

Together, these two directions---machine-learning tools that design magnonic devices, and magnonic devices that themselves compute---mark a convergence of magnonics and artificial intelligence that we term \emph{AI magnonics} (Fig.~\ref{fig:ai-magnonics});  we define the term more fully in Sec.~\ref{sec:universal-device}.

\subsection{Open questions and fundamental limits}
\label{sec:open-questions}

The rapid progress surveyed above should not obscure how much remains genuinely unknown, and we believe the field is best served by stating these unknowns plainly.
A central open question concerns scale: how does the complexity of a realisable function grow as the design space is enlarged from $N$ to $2N$ or $N^2$ parameters?
Cascading inverse-designed blocks has yielded a magnonic half adder~\cite{Chen2025VCMA}, but does the same paradigm extend to substantially larger circuits---a multi-bit adder, for instance---or does performance degrade faster than added degrees of freedom can compensate (Sec.~\ref{sec:parametric-amp})?

Even this is subtler than it sounds, because the number of degrees of freedom is neither measured consistently nor always indicative of the same thing.
Reported counts span more than four orders of magnitude---from five geometry parameters~\cite{Yan2022} to of order $10^5$ material cells~\cite{Kiechle2022}---yet are not directly comparable: the count of independent parameters differs from the number of reachable configurations (the 49-loop field device spans up to $\sim\!10^{162}$ states, of which $\sim\!10^{87}$ were used in practice~\cite{Zenbaa2025NatElec, Zenbaa2025SciAdv}).
A discretisation grid bounds the configuration space in either case, but the meaningful count depends on the parameterisation: each cell may be an independent parameter (pixel topologies~\cite{Wang2021, Chen2025VCMA}) or merely the resolution onto which a lower-dimensional boundary representation is rendered (level-set methods~\cite{Voronov2025}).

A related question concerns footprint.
In CMOS, processing capacity scales with area: a full adder occupies roughly twice the footprint of a half adder.
Inverse design promises a different trade-off, accommodating greater functional complexity by making a structure of fixed size internally more intricate rather than larger~\cite{Zenbaa2025NatElec}---but how far can this intricacy be pushed?
Diffraction sets the bound: the smallest usable feature scales with the spin-wave wavelength.
Here magnonics is particularly well placed: spin-wave wavelengths have been demonstrated down to about 50\,nm~\cite{Liu2018} and are in principle tunable to their fundamental limit, the lattice constant, where frequencies rise into the THz range.

Further questions cut across the frontiers above.
Can stochastic nonlinearities be exploited by design rather than opportunistically (Sec.~\ref{sec:nonlinear-dof})?
Can the five design variables and the gain distribution (Sec.~\ref{sec:parametric-amp}) be co-optimised within a single optimisation loop?
And, most fundamentally, is inverse design truly universal, or do certain classes of functionality lie beyond its reach?
We regard these less as caveats than as important research questions of the coming years.

\subsection{The universal magnonic device}
\label{sec:universal-device}

A striking feature of inverse design is its universality as a method.
The same procedure---specify a target as an objective function, then let an optimiser find the structure---has already produced functionally unrelated devices spanning wave optics, RF signal routing and filtering, Boolean logic, and neuromorphic pattern recognition (Table~\ref{tab:demonstrations}); from one case to the next, only the objective function is swapped.
The same universality extends across physical domains: magnon-phonon coupling for surface acoustic wave devices, magnon-photon interfaces for microwave-to-optical transduction, and magnon-superconductor hybrid systems are regimes where intuitive design is particularly difficult, if not impossible, and inverse design may prove most valuable.
It also invites standardisation: a library of validated inverse-design building blocks---routers, filters, logic gates---could enable modular magnonic circuit design, and common benchmark problems would facilitate fair comparison.
Such problems necessarily depend on the target functionality; a two-frequency demultiplexer or an all-magnonic half-adder judged on cascadability would be suitable first candidates. The quantities that make results comparable are more general: insertion loss, isolation, bandwidth, the on--off ratio for logic operations, robustness, and footprint in units of the spin-wave wavelength.

More broadly, the convergence of Sec.~\ref{sec:self-adapting}---\emph{AI magnonics} (Fig.~\ref{fig:ai-magnonics})---points beyond the individual frontiers discussed above: a paradigm in which machine-learning algorithms design magnonic devices that themselves perform neuromorphic computation, closing the loop between artificial intelligence and the physical platform it optimises.
What the term names is this development as a whole, not a new subfield. It does not replace the established names for this work, such as neuromorphic magnonics, magnonic computing, or physical reservoir computing.
This convergence culminates in the long-range vision of a universal magnonic device.
Whereas modular circuits assemble many specialised blocks, the universal device is a single reconfigurable platform, reprogrammable for any functionality via software-defined field landscapes or non-volatile magnetic textures at the nanoscale.
In practice, such a platform would be trained once for a whole set of functionalities and shipped with it: switching then reduces to reloading a stored configuration, trading the slow one-time training (minutes to hours) for reconfiguration on nanosecond timescales~\cite{Zenbaa2025NatElec}.

The trajectory of mainstream artificial intelligence offers a suggestive analogy: many capabilities of large language models---multilingual fluency among them---were never explicitly programmed, but emerged once a general machine-learning infrastructure was scaled.
Inverse-design magnonics is built on that same infrastructure (Sec.~\ref{sec:diff-solvers}), now applied to a physical spin-wave medium rather than a purely digital model.
We anticipate that this convergence could transform spin-wave science from a field of device demonstrations into a systematic engineering discipline.


\section*{Acknowledgments}
The authors acknowledge fruitful discussions with members of the Physics of Functional Materials and the Nanomagnetism and Magnonics groups at the University of Vienna.

\section*{Funding}
This research was funded in whole or in part by the Austrian Science Fund (FWF) [grant DOIs 10.55776/PIN1434524 and 10.55776/PAT3864023]. For open access purposes, the authors have applied a CC BY public copyright license to any author-accepted manuscript version arising from this submission.

\section*{Conflict of interest}
The authors declare no conflict of interest.

\section*{Author contributions}
F.V.: conceptualisation, visualisation, writing -- original draft. F.B.: writing -- review \& editing. C.A.: writing -- review \& editing. D.S.: funding acquisition, supervision, writing -- review \& editing. A.C.: conceptualisation, funding acquisition, supervision, writing -- review \& editing. All authors discussed the content and contributed to revising the manuscript.

\section*{Data availability}
No new data were created or analysed in this study.

\bibliographystyle{ieeetr}
\bibliography{InverseDesign}

\end{document}